\documentstyle[epsf,12pt,epsfig]{article}  
\begin{document}


\begin{titlepage}

\pagenumbering{arabic}
\vspace*{-1.5cm}
\begin{tabular*}{15.cm}{c}
{\large{EUROPEAN ORGANIZATION FOR PARTICLE PHYSICS}  }
\end{tabular*}
\vspace*{2cm}
\begin{tabular*}{15.cm}{l@{\extracolsep{\fill}}r}
 & \\
 & CERN-OPEN 98-020  
\\
& 
2 August, 1998
\\
&\\ 
\end{tabular*}
\vspace*{2.cm}
\begin{center}
\Large 
{\bf 
Estimating the Parameters of Bose-Einstein Correlations from the
Two-Particle Correlation Function 
in Multihadronic Final States
}\\
\vspace*{0.6cm}
   {\normalsize 
   A.~De Angelis\footnote{CERN EP Division, CH-1211 Geneva 23, Switzerland}
    and 
    L.~Vitale\footnote{Dipartimento 
  di Fisica, Universit\`a di Trieste and
     INFN, Via A. Valerio 2, I-34127 Trieste, Italy}
}
\vspace*{2mm}
\vspace*{1.5cm}
\end{center}
\vspace*{2cm}
\begin{abstract}
\noindent
To estimate the strength of the Bose-Einstein correlations and the
radius of the hadronization region in multiparticle production, 
the two-particle correlation functions $R$ for identical pairs
is adjusted to a parametric function describing the enhancement at small
momentum differences. 
This is usually done by
means of a binned uncorrelated least squares fit. 
This article demonstrates that this procedure 
underestimates the statistical errors.
A recipe is given to construct from the data the covariance matrix between the 
bins of the histogram of the two-particle correlation function.\\
{\em PACS codes: 07.05.K, 05.30. 
}
\end{abstract}
\clearpage

\end{titlepage}

\setcounter{page}{1}    

\section{Introduction}

An enhancement in the production of pairs of pions of the same charge and
similar momenta
produced in high energy collisions was first observed in
antiproton annihilations
and attributed to Bose-Einstein statistics appropriate to identical
pion pairs \cite{gol}.

Bose-Einstein Correlations (BEC) between pion pairs can be used to study the
space-time structure of the hadronization source \cite{kop}. 
This has been done for hadron-hadron,
heavy ion, muon-hadron and
$e^{+}e^{-}$ collisions (see for example 
Ref. \cite{rev} for reviews). 

The precise measurement of BEC parameters is especially important for
 LEP 2 physics. 
Interference due to 
Bose-Einstein correlations 
in hadronic decays of WW pairs has been discussed on a theoretical basis,
in the framework of the measurement of the W mass \cite{thww}:
this interference could induce a systematic uncertainty on the W
mass measurement in the 4-jet mode which is of the order of 40 MeV, i.e.,
comparable with the expected accuracy of the measurement. 
For a review of measurements of BEC in WW pairs, see 
for example \cite{deafra}.

\section{Determination of BEC parameters}

To study the enhanced probability for the emission of two identical bosons,
the correlation function $R$ is used as a probe. For pairs of particles, 
it is defined as 
\begin{equation}
R(p_{1},p_{2}) = \frac{P(p_{1},p_{2})}{P_{0}(p_{1},p_{2})} \, ,
\end{equation}
where $P(p_{1},p_{2})$ is the two-particle probability density, subject to
Bose-Einstein symmetrization, $p_{i}$ is the four-momentum of
particle $i$, and $P_{0}(p_{1},p_{2})$ is a
reference two-particle distribution which, 
ideally, resembles $P(p_{1},p_{2})$ in all respects, apart from the lack
of Bose-Einstein symmetrization. 

If $d(x)$ is the space-time distribution of the source, $R(p_{1},p_{2})$ 
takes the form
\begin{eqnarray*}
R(p_{1},p_{2})=1+|G[d(x)]|^{2} \, ,
\end{eqnarray*}
where $G[d(x)]=\int{d(x)e^{-\imath(p_{1}-p_{2})\cdot x \,} dx}$ 
is the Fourier transform of $d(x)$. 
Thus, by studying the correlations between the momenta of pion pairs,
one can determine the distribution of the points of origin of the pions.
Experimentally, 
the effect is often described in 
terms of the Lorentz-invariant variable $Q$, defined by 
$Q^{2}=(p_1-p_2)^2=M^{2}(\pi \pi)-4m^{2}_{\pi}$,
where $M$ is the invariant mass of the two pions.
The correlation function can then be written as
\begin{equation}
R(Q) = \frac{P(Q)}{P_{0}(Q)},
\end{equation}
which is usually parametrized by the function  
\begin{equation}\label{beq}
R(Q)= N \left( 1 + \lambda e^{-r^2Q^2} \right) \, .
\end{equation}
In the above equation, the pion source is spherically symmetric and gaussian, 
the parameter $r$ gives the RMS source radius, $\lambda$ is
the strength of the correlation between the pions and $N$ an overall 
normalization factor. 
The data from $e^+e^-$ annihilations from PEP energies to LEP show values of 
$r$ around 0.6 fm; the value of $\lambda$ strongly depends on the analysis 
technique.

It can be understood from what said above 
that the main problems in the study of BEC 
are given by a good choice of the reference sample, 
and by the definition of the normalization $N$. 

Most of the experimental studies are done inclusively, using all
the observed charged tracks measured.
Electrons, charged kaons and protons do not correlate with pions and 
clearly reduce the experimental correlation function.
Another reduction of $R(Q)$ is due to non-prompt pions, i.e., 
pions from decays of particles with lifetime larger than the
hadronization scale of around 1 fm/$c$ 
(like K$^0$, $\Lambda$ and $b$ and $c$ hadrons).
These pions are not expected to correlate with those from the primary 
hadronization.

\section{Analysis}

The analysis performed in this paper 
was based on the JETSET simulation \cite{lund}.
JETSET has shown to reproduce well the experimental 
inclusive distributions measured by LEP \cite{tuning}.

BEC were described by the
Bose-Einstein simulation algorithm LUBOEI, fully integrated in
the JETSET simulation.  
The values generated for the pion momenta 
are modified by this algorithm, which reduces the differences for 
pairs of like-sign particles.
This code has been shown \cite{delbe} to reproduce well the 
two particle correlation functions measured in Z decays  
if Bose-Einstein correlations are switched on 
with a Gaussian parametrization for pions that are 
produced either promptly 
or as decay products of short-lived resonances 
(resonances with  lifetime longer than the K$^*(890)$ lifetime  
were considered long-lived)
and if the parameter values  $\lambda=1$ and $r=0.5$~fm are
used as input\footnote{The measured 
values of the parameters with a mixing reference sample 
for such ``direct'' pions in Z decays
were $\lambda=1.06 \pm 0.17$, $r=0.49 \pm 0.05$ fm \cite{delbe}.}.
The value $\lambda$=1 for direct pions corresponds to 
$\lambda\sim0.35$ for all pions or 
$\lambda\sim0.25$ for all particles~\cite{delbe}.
The fitted value
of the radius $r$ depends on the choice of 
the reference sample. Using a Monte Carlo
reference sample changes typically
$r$ from the input value of 0.5 fm to a 20\% higher value.

The use of simulated samples gives the possibility of
knowing the normalization and of defining an unbiased reference sample 
simply by
``switching off'' the Bose-Einstein correlations.

\section{The naive approach to the fit gives wrong results}

We choose as a case study the hadronic decay of WW pairs 
with full Bose-Einstein effect and
the BEC parameters set to $\lambda=1$ and $r=0.5$~fm respectively.
400 samples of 2,000 events each 
were simulated. For the reference sample $P_0$, 
40,000 events were simulated without BEC, in such a way that the error
on $P_0$  gave a negligible contribution to the error on $R$.  

For each sample a histogram of the correlation function was built,
using 40 bins of 50 MeV each from 0 to 2 GeV.
The normalization factor $N$ was fixed by imposing that the average value
of $R$ between 1 GeV and 2 GeV was equal to unity.

For each of the 400 samples 
we performed a $\chi^2$ fit to
the form (\ref{beq}). 
The average values of $\lambda$ and $r$ from the 400 fits were:
\begin{eqnarray*}
\lambda & = & 0.509 \pm 0.016 \, ,\\
r & = & 0.569 \pm 0.012~{\mathrm{fm}}
\end{eqnarray*}
(the statistical error corresponds to the average of the errors).

\begin{figure}[hbt]
\begin{center}
\mbox{\epsfxsize16.0cm\epsffile{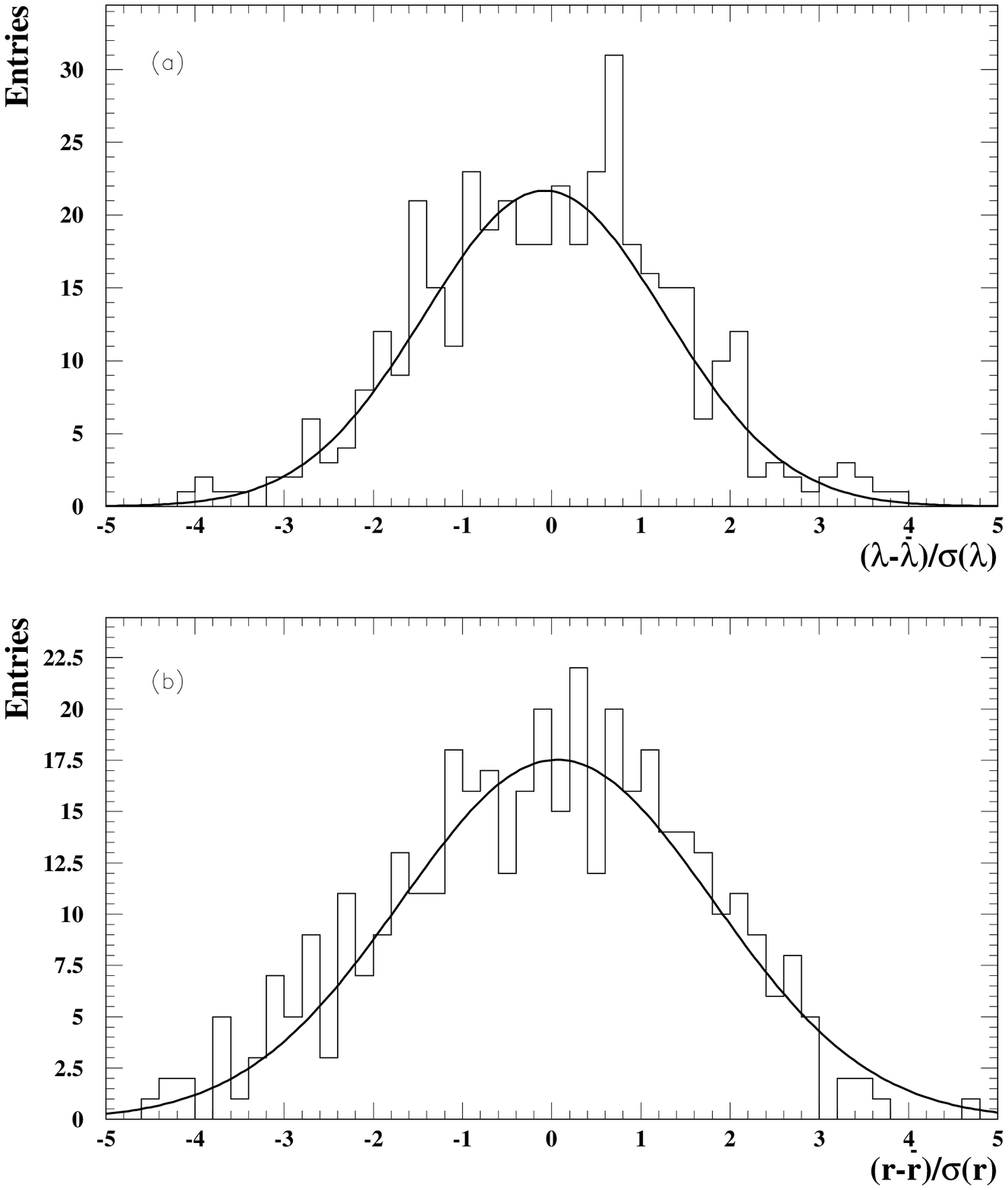}}
\end{center}
\caption{\label{f1} 
(a) Pull function for the fitted parameter $\lambda$ 
using a binned uncorrelated least squares fit 
in the simulated samples.
A Gaussian fit is superimposed as a solid line.
(b) Same as (a) but for the parameter $r$. 
}
\end{figure}

The ``pull'' of the fitted values of $\lambda$ and $r$
is shown in Figure \ref{f1}.
A Gaussian fit gives
$\sigma_{(\lambda-<\lambda>)/\sigma_\lambda} = 1.35$ with $\chi^2$/DoF = 35/37,
and $\sigma_{(r-<r>)/\sigma_r} = 1.75$ with $\chi^2$/DoF = 30/38.

The error on $r$ is thus underestimated by a factor 1.75
while the error on $\lambda$ is underestimated by a factor 1.35.
The fact that these factors are different indicates that
the ``naive'' fitting procedure described in this section could give
possible biases in the determination of the parameters. 

When WW pairs in which 
one W decays hadronically and the other leptonically are simulated,
the pulls are smaller (1.30 for $r$ and 
1.24 for $\lambda$). This can be explained by the
lower multiplicity, which gives smaller correlations.
The case of the Z particle is similar to the semileptonic W.

\section{Improved technique for the evaluation}

The presence of bin-to-bin correlations in $R$
is unavoidable: if there are, say, $M$ positive tracks,
the same positive 
track enters $(M-1)$ times in the two-particle density $P$.
We build the generic nondiagonal term $C_{ij}$ $(i \neq j)$ of the 
covariance matrix 
by assuming that it is given by the number of
times that a track entering in the $i$th-bin enters also in the $j$th-bin.

However, this is not the only statistical correlation effect, 
since the same track can also enter several times in the same bin.
The latter effect can be accounted for by rescaling the error of each bin, 
$\sigma(b_i)$, to the expected relative 
error for the true independent counts.
If we call $r_i$ the number of extra counts due to tracks already present 
once in the bin $i$, then the number of independent counts $d_i$ for the bin i
is $d_i=b_i-r_i$ and
\begin{equation} \label{miodia}
\frac{\sigma(b_i)}{b_i} = \frac{1}{\sqrt{d_i}}  \, .
\end{equation}
The diagonal terms $C_{ii}$ of the 
covariance matrix are
assumed to be the squares of $\sigma(b_i)$.

\begin{figure}[hbt]
\begin{center}
\mbox{\epsfysize20.0cm\epsffile{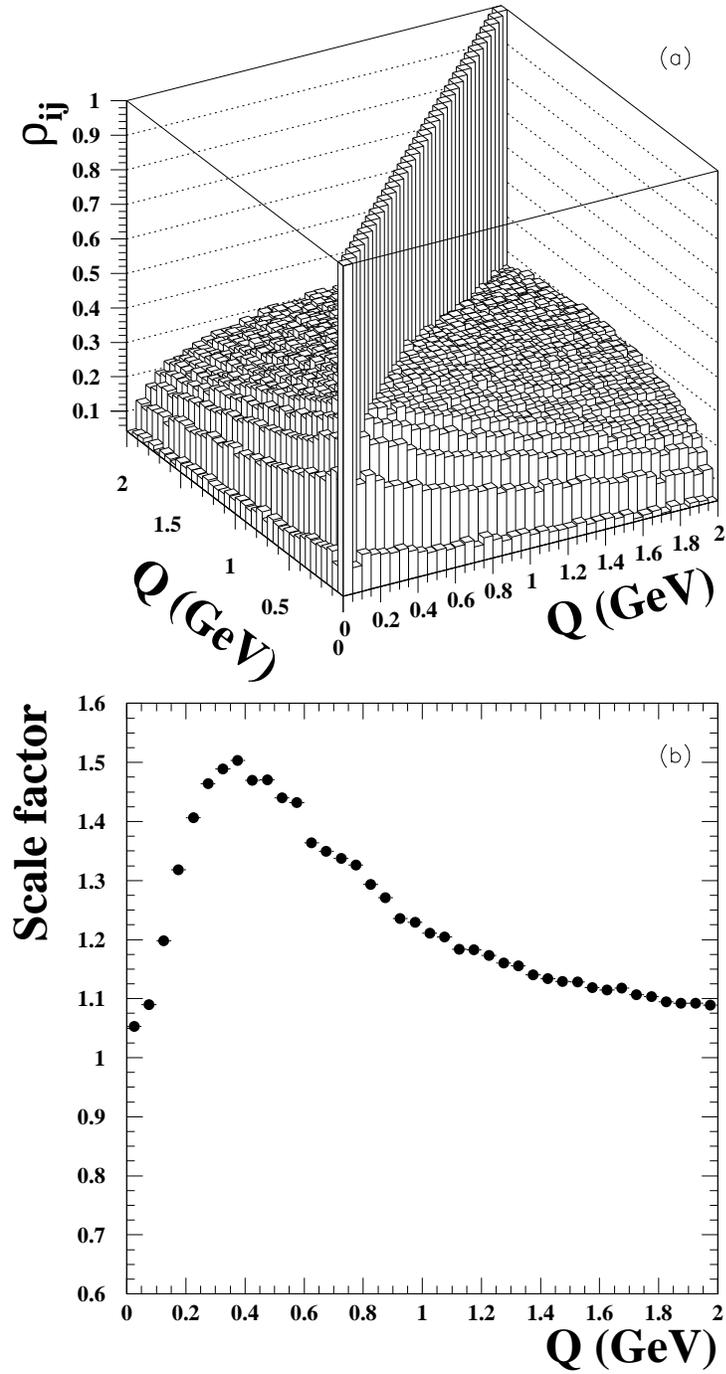}}
\end{center}
\caption{\label{f2} 
(a) Correlation matrix (see text).
(b) Ratio between the errors on the bin contents 
estimated with the procedure described in this paper and the
errors computed from the naive approach. 
}
\end{figure}

The correlation matrix $\rho_{ij} = C_{ij}/(\sigma_i\sigma_j)$
looks like in Figure \ref{f2}a.
In Figure \ref{f2}b, the
ratio between the errors on the bin contents 
estimated from Eq. (\ref{miodia}) and the
square root of the number of entries $b_i$. 

For each of the 400 samples 
we performed a $\chi^2$ fit to
the form (\ref{beq}), this time by using the 
covariance matrix described above. 
The average values of $\lambda$ and $r$ from the 400 fits were:
\begin{eqnarray*}
\lambda & = & 0.499 \pm 0.021 \, ,\\
r & = & 0.581 \pm 0.017~{\mathrm{fm}}
\end{eqnarray*}
(the statistical error corresponds to the average of the errors).

\begin{figure}[hbt]
\begin{center}
\mbox{\epsfxsize16.0cm\epsffile{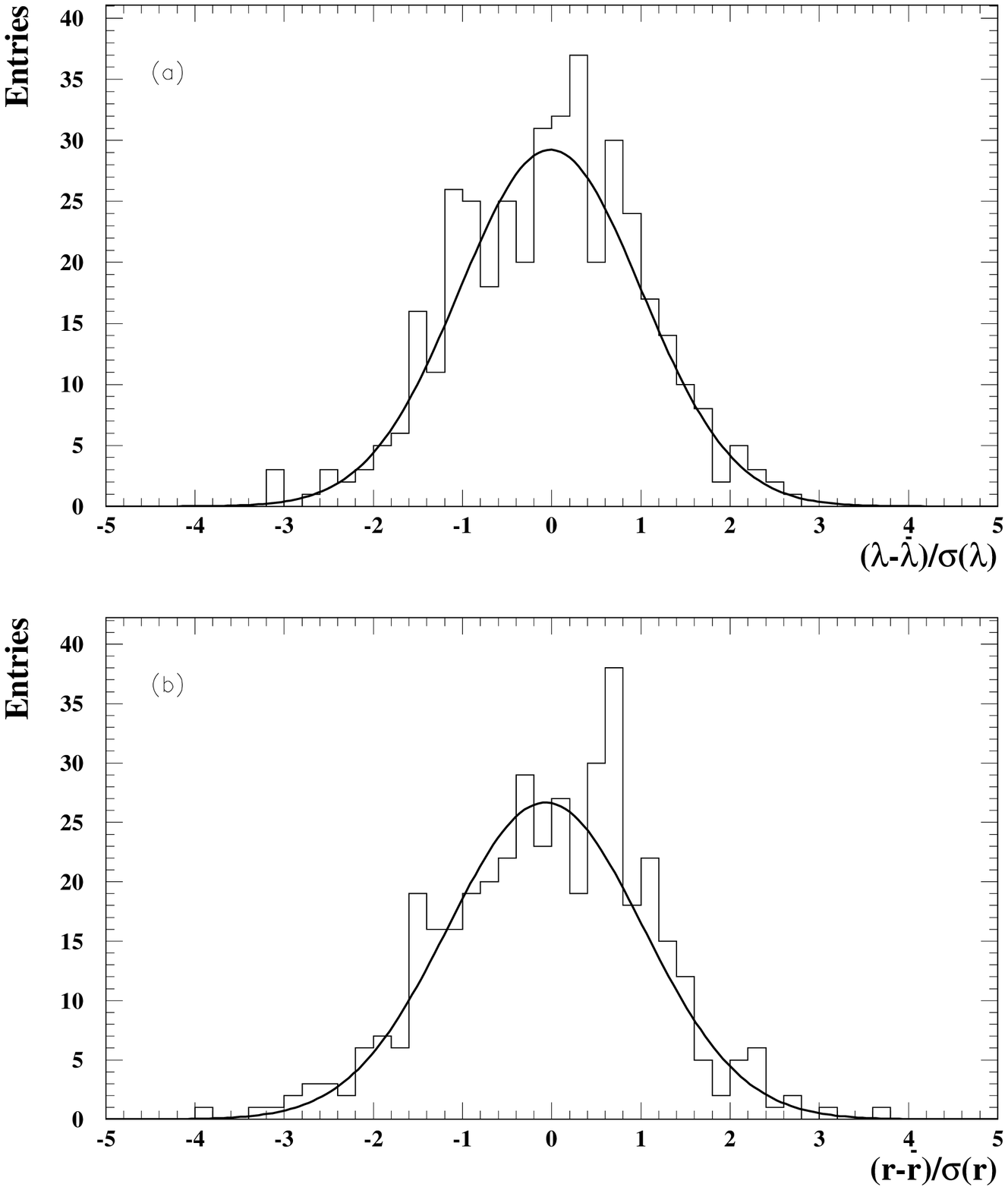}}
\end{center}
\caption{\label{f3} 
(a) Pull function for the fitted parameter $\lambda$  
in the simulated samples
using the technique for the error estimate described in this
paper.
A Gaussian fit is superimposed as a solid line.
(b) Same as (a) but for the parameter $r$. 
}
\end{figure}

The ``pull'' of the fitted values of $\lambda$ and $r$
is shown in Figure \ref{f3}.
A Gaussian fit gives
$\sigma_{(\lambda-<\lambda>)/\sigma_\lambda} = 1.02$ with $\chi^2$/DoF = 28/26,
and $\sigma_{(r-<r>)/\sigma_r} = 1.09$ with $\chi^2$/DoF = 35/31.

The error on $r$ and $\lambda$ 
are thus correctly estimated (at better than 10\%). A residual discrepancy
can be attributed to the fact that, due to a $Q$-dependence of the fraction
of particles which do non display BEC, the fitting function (\ref{beq})
does not describe the data perfectly.

\begin{figure}[hbt]
\begin{center}
\mbox{\epsfxsize16.0cm\epsffile{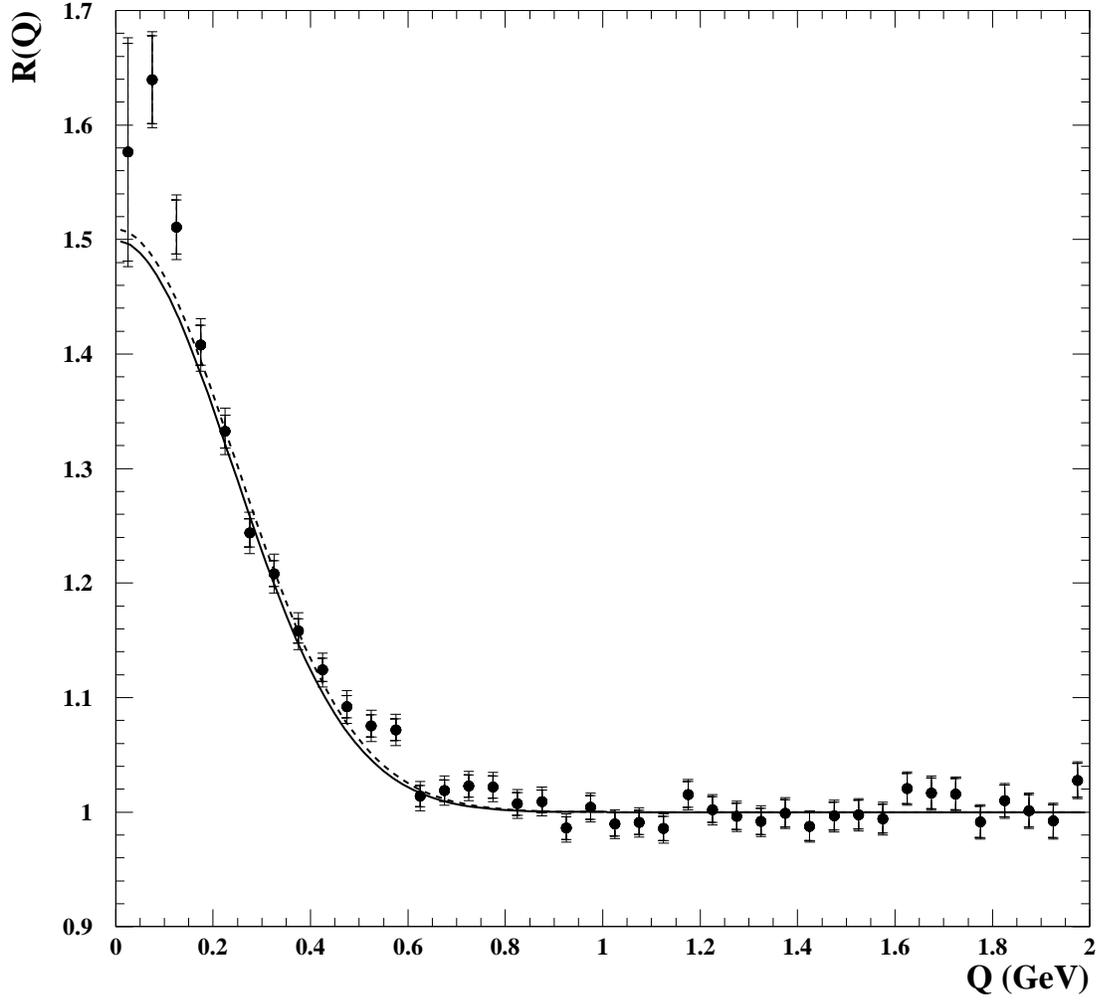}}
\end{center}
\caption{\label{f4} 
The correlation function $R(Q)$ for like-sign pairs (closed circles)
in one of the simulated
samples. The error bars shown are the square roots
of the diagonal elements of the covariance matrix.
The inner error bars correspond to the error 
computed from the naive approach.
The dashed line shows the fit to
Eq. (\ref{beq}) where $\lambda$ and $r$ are the
averages of the fitted values using the
naive approach. The solid line corresponds to the approach proposed in
this article.
}
\end{figure}

As an example, in Figure \ref{f4} 
the correlation functions $R(Q)$ for like-sign pairs (closed circles)
in one of the 400 simulated
samples is shown together with 
the diagonal errors from the naive approach and 
from the approach proposed in this paper, 
and with the fitting function 
corresponding to the average results of the fit in the
two approaches. 

\section{Conclusions}

Using a binned uncorrelated least squares fit to evaluate the parameters 
$\lambda$ and $r$ of Bose-Einstein correlations  
underestimates the statistical errors and might bias the result.

The error underestimate has been shown to be of the order of 30\% in the
case of single W and Z and of about 50\% in the case of the
hadronic decay of W pairs. 
A recipe is given to construct from the data 
a heuristic covariance matrix between the 
bins of the histogram of the two-particle correlation function.

\subsection*{Acknowledgements}
We are grateful to G. Della Ricca for useful discussions.

\newpage
\setlength{\unitlength}{0.7mm}


\newpage

\newpage

\newpage

\end{document}